\documentclass[twocolumn,showpacs,aps]{revtex4}

\usepackage[dvips,final]{graphics}
\usepackage{textcomp}
\usepackage{subeqnarray}

\begin{document}

\draft

\newcommand{\no}{\nonumber}
\newcommand{\etal}{{\em et~al }}
\newcommand{\ie}{{\it i.e.\/}\ }

\title{\center{Non-gaussian statistics from individual pulses of squeezed light}}

\author{J\'er\^ome Wenger, Rosa Tualle-Brouri and Philippe Grangier}
\affiliation{Laboratoire Charles Fabry de l'Institut d'Optique,
CNRS UMR 8501, F-91403 Orsay, France. {\rm e-mail :
jerome.wenger@iota.u-psud.fr}}

\begin{abstract}
We describe the observation of a ``degaussification" protocol that
maps individual pulses of squeezed light onto non-Gaussian states.
This effect is obtained by sending a small fraction of the
squeezed vacuum beam onto an avalanche photodiode, and by
conditioning the single-shot homodyne detection of the remaining
state upon the photon-counting events. The experimental data
provides a clear evidence of phase-dependent non-Gaussian
statistics. This protocol is closely related to the first step of
an entanglement distillation procedure for continuous variables.
\end{abstract}
\maketitle

Researches on novel schemes to perform quantum key distribution
(QKD) are presently very active. In that field, lots of interest
has arisen recently on the use of quantum continuous variables
(QCV). For instance novel QKD schemes using the quadrature
components of amplitude and phase modulated coherent states have
been recently proposed \cite{prl} and experimentally demonstrated
\cite{GVAWBCG03}. It has been shown that such coherent state
protocols are secure against individual gaussian attacks for any
value of the line transmission \cite{GVAWBCG03,qic}, and actually
more general proofs are presently under study \cite{gc,IVAC}.

An important practical advantage of coherent states QKD is that it
can in principle reach very high secret bit rates
\cite{GVAWBCG03}. However, even in the best possible case,
coherent states QKD will not do much better than photon-counting
QKD  \cite{gisin} in terms of absolute distance, because of the
exponential attenuation in optical fibers~: at some point which is
now somewhere between 10 and 100 km, one hits a limit where the
transmitted secret data gets buried into errors of various
origins, that range from detectors dark counts to imperfect data
processing.

In order to qualitatively improve the situation, \ie to go much
beyond the attenuation length of a strand of fiber, a major
challenge is to implement quantum repeaters \cite{qr}, based upon
entanglement distillation and (most likely) quantum memories.
Ultimately, the secret qubits would be simply teleported to a
remote place, with which shared entanglement has been established
\cite{bennett}. Looking now at entanglement distillation for QCV,
a difficulty appears quickly~: most (if not all) QCV transmissions
so far are using light beams with gaussian statistics. However, it
has been shown that it is not possible to distillate entanglement
from a gaussian input to a gaussian output by gaussian means
\cite{eisert1,cirac}. One has to jump ``outside" the gaussian
domain, though it is possible to reach it back at the end, at
least in an approximate way \cite{eisert2}.

In this letter, we experimentally implement a procedure which we
call ``degaussification", that maps short pulses of squeezed light
onto non-Gaussian states. This protocol is based upon a
post-selection triggered by a photon-counting event and uses only
simple linear optical elements. Extending this procedure to
entangled EPR beams -which is fairly simple in principle- provides
the first step of an entanglement distillation procedure as
proposed in ref. \cite{eisert2}.

\begin{figure}[t]
\center %\vskip 3.8cm
\includegraphics{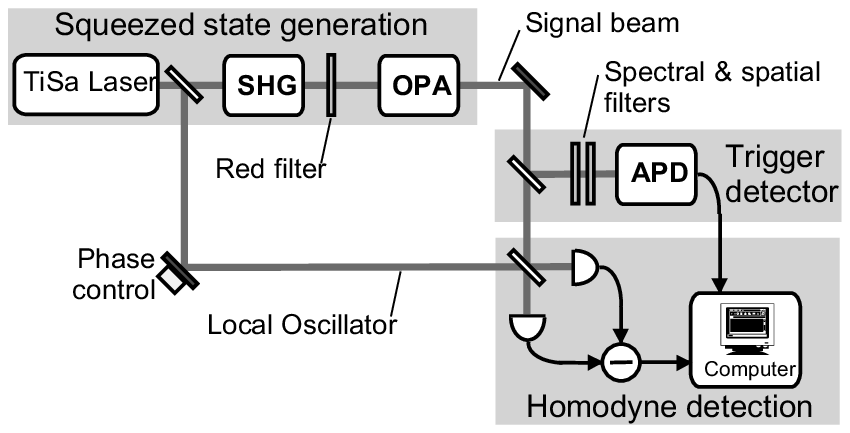}
\caption{Simplified experimental setup} \label{Expsetup}
\end{figure}

The experimental scheme is presented on Fig.~\ref{Expsetup}. The
initial pulses are obtained from a titanium-sapphire laser
(Tiger-CD, Time-Bandwidth Products), delivering nearly
Fourier-transform limited pulses at 850 nm, with a duration of 150
fs, an energy of 40 nJ, and a repetition rate of 790 kHz. These
pulses are frequency doubled in a single pass through a thin (100
$\mu$m) crystal of potassium niobate (KNbO$_3$), cut and
temperature-tuned for non-critical type-I phase-matching. The
second harmonic power is large enough to obtain a significant
single-pass parametric gain ($\sim$ 3 dB) in a similar KNbO$_3$
crystal used in a type-I spatially degenerate configuration.

Given this relatively high gain, ``real" squeezed states are
actually produced, not only parametric pairs. Therefore, higher
order terms (beyond pair production) have explicitly to be
included in the analysis as they play an essential role to
understand the phase-dependence of the data. The detection scheme
follows the basic idea of a pulsed squeezed light experiment
\cite{laporta}, with two important differences~:

(i) All processing is done in the time domain, not in the
frequency domain. For each incoming pulse, the balanced homodyne
detection samples one value of the signal quadrature in phase with
the local oscillator beam \cite{GVAWBCG03}. It is then possible to
reconstruct the full statistics of the signal pulses. The
histograms presented below are obtained from these individual
pulse data.

(ii) A small fraction ($R=0.115$) of the squeezed vacuum beam is
taken out from the homodyne detection channel. These trigger
photons then pass through a spatial filter (made of two
Fourier-conjugated pinholes) and a 3 nm spectral filter centered
at the laser wavelength, before being detected by a silicon
avalanche photodiode (APD). The detection click is registered
simultaneously with the homodyne signal, and can be used to
post-select homodyne events. As we will show, this selection
provides directly non-gaussian statistics.

\begin{figure}[t]
\center %\vskip 3.5cm
\includegraphics{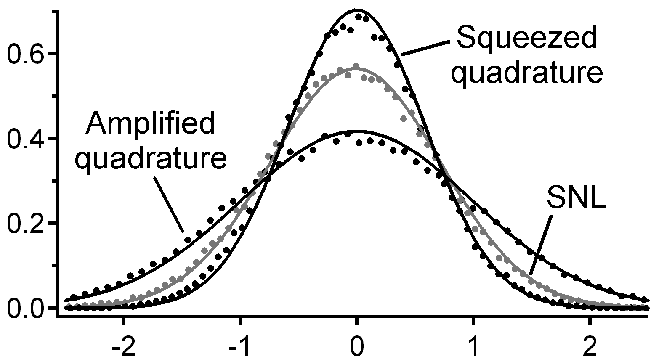}
\caption{Normalized probability distribution for the
(unconditioned) squeezed vacuum state, obtained from the pulsed
homodyne detection. The squeezed quadrature variance is 1.75 dB
below SNL, while the amplified quadrature variance is 3.1 dB
above. The SNL curve corresponds to the vacuum state, where the
shot noise variance is taken equal to 1/2.} \label{squeez}
\end{figure}

The unconditioned distributions corresponding to the squeezed and
anti-squeezed quadratures, and to the vacuum noise are plotted on
fig.~\ref{squeez}. More experimental details about the squeezed
states generation will be given in another publication. The
measured squeezing variance (with no correction) is 1.75 dB below
the shot noise level (SNL), in good agreement with the measured
deamplification of a probe beam (0.50 or 3 dB) and our evaluation
of the overall detection efficiency $\eta_{tot} = \eta \; (1-R) =
0.66$. Here $1-R=0.885$ is the transmission of the conditioning
beamsplitter, and $\eta=0.75$ is the homodyne detection efficiency
(see details below). As it can be seen on  fig.~\ref{squeez}, the
experimental data for both quadratures is correctly fitted by
assuming a single-mode parametric gain $\exp(\pm 2 s)$ with $s =
0.43$, together with the above efficiency $\eta_{tot}$. We note
however that the deamplification gain of the probe beam does not
correspond exactly to the inverse of the amplification, due to
gain-induced-diffraction which distorts the probe phase fronts
\cite{Gid}. Since such multimode effects remain reasonably small
in our experimental conditions, we will use the single parameter
$s$ to describe parametric amplification and deamplification.

\begin{figure}[t]
\center %\vskip 7cm
\includegraphics{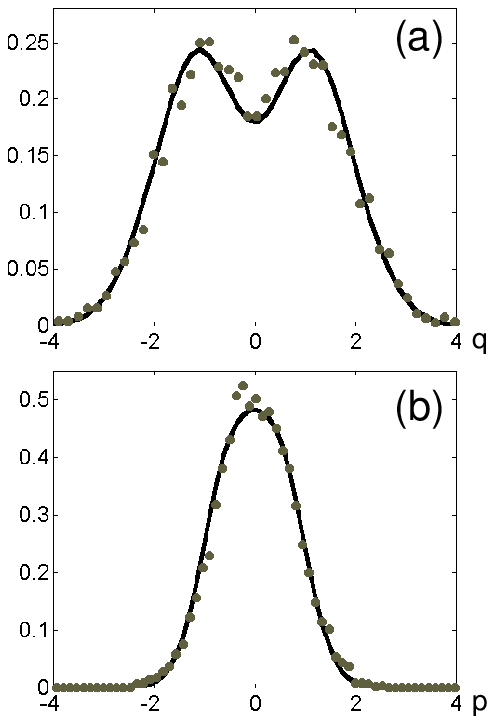}
\caption{Experimental (dots) and theoretical (line) quadrature
distribution of the post-selected homodyne measurements for the
amplified quadrature (a) and the squeezed one (b), normalized as
in fig.~\ref{squeez}. Parameters used in the calculation are
$s=0.43$, $R=0.115$, $\eta=0.75$ and $\xi=0.7$ .}
\label{squeezcond}
\end{figure}

Fig.~\ref{squeezcond} displays the post-selected output of the
homodyne detection resulting from the degaussification protocol,
showing a clear dip in the centre of the amplified quadrature
distribution. The theoretical curves represented on the same
figure are obtained from a simple single-mode model detailed
below. This model takes into account the measured parametric gain,
together with various experimental imperfections (losses,
imperfect mode-matching, electronic noise, dark counts and modal
purity, see below for details), and it is clearly in good
agreement with the experimental data.

The origin of the  observed effect can be analyzed in different
ways. A first insight can be obtained by considering the homodyne
detection of a conditional single photon state, observed in
\cite{ek}. In this experiment, the authors separate the two
photons from a parametric pair, and one of them is used as a
trigger on a photon counter, while the other one is sent to an
homodyne detection. In the ideal case, such an experiment would
measure the probability density $P(x)$ of the $n=1$ Fock state,
which is non-gaussian since $P(0)=0$. Though this experiment
provides a first idea of the origin of the non-gaussian features,
it is not enough to explain our observations. Actually, we see
phase-dependent effects (while a $n=1$ Fock state is
phase-independent), and in our set-up there is no explicit
separation of the photon pair.

\begin{figure}[t]
\center %\vskip 3.2cm
\includegraphics{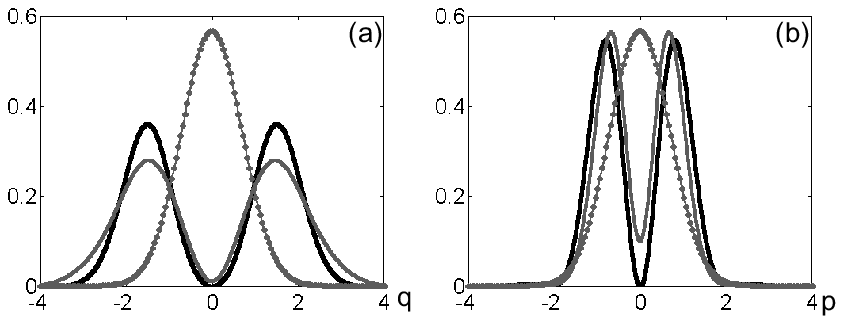}
\caption{Phase-dependent quadrature distributions of the
conditioned homodyne measurements, together with the vacuum
reference (line and dots). The thick solid line is obtained from
eq. (\ref{condstate}) with $R=0.01$. The thin gray line is
obtained from the complete calculation and $R=0.115$. Fig.~(a)
corresponds to the amplified quadrature while fig.~(b) shows the
squeezed one. The squeezing parameter is $s = 0.43$, and perfect
single mode detection efficiency has been assumed.}
\label{condtheo}
\end{figure}

We have carried out a calculation taking into account the
expansion of the squeezed state in a Fock state basis, including
terms up to $n=10$, which is enough for our degree of squeezing.
The calculation is done for an arbitrary value of the conditioning
beamsplitter reflectivity, and takes into account the various
imperfections of the experiment. This calculation is
straightforward but tedious, and one can actually get a good
physical insight by considering the restricted simple case of an
expansion of the squeezed vacuum up to $n=4$, and a beamsplitter
reflectivity $r=\sqrt{R}\ll1$. The squeezed vacuum
$\mid\Psi_s\rangle$ can then be written as~:
\begin{equation}
\mid\Psi_s\rangle = \alpha \mid0\rangle+ \beta \mid2\rangle+
\gamma \mid4\rangle
\end{equation}
With our degree of squeezing $s=0.43$, one has $\alpha=0.96$,
$\beta=0.27$ and $\gamma=0.10$ \cite{ulf}. This state then gets
mixed with the vacuum at the beamsplitter, resulting in a two-mode
entangled squeezed state. Denoting as $r,t$ the reflectivity and
transmittance of the beamsplitter ($r^2+t^2=1$), the output state
is~:
\begin{eqnarray}
\mid \Psi_{s,out}\rangle =
\left(\alpha\mid0\rangle_1+t^2\beta\mid2\rangle_1+t^4\gamma\mid4\rangle_1\right)\mid0\rangle_2
\nonumber \\
+
\left(\sqrt{2}rt\beta\mid1\rangle_1+2rt^3\gamma\mid3\rangle_1\right)\mid1\rangle_2+O(2)
\end{eqnarray}
where $\mid.\rangle_1$ denotes the state sent to the homodyne
detection, while $\mid.\rangle_2$ stands for the state sent to the
APD. The term $O(2)$ denotes Fock state terms higher than 1 on the
APD beam, which will be neglected in this simplified calculation,
given our assumption $r\ll1$. Finally, post-triggering on the APD
photon-counting events reduces the state detected by the homodyne
detection to~:
\begin{equation} \label{condstate}
\mid\Psi_{cond}\rangle \propto \beta \mid1\rangle+ \sqrt{2}\gamma
\; t^2 \mid3\rangle
\end{equation}
The prediction of this calculation is shown on
fig.~\ref{condtheo}. As it could be expected, we do obtain
phase-dependent non-gaussian statistics. These features are
related to high order terms beyond pair production which play an
essential role in our analysis.

In this simplified calculation we have assumed $ r <<1$, and the
predicted dip in the centre of the probability distribution goes
down to zero. When the beamsplitter reflectivity is increased,
Fock state terms with $n>1$ may no longer be neglected on the APD
beam, and the central dip has a non-zero value. Strictly speaking,
this is not an experimental imperfection, but an intrinsic feature
of the conditioned state for larger $R$, which clearly appears on
the result of the full calculation also displayed on Fig.~4.

In order to characterize experimental imperfections, let us
emphasize that the homodyne detection and the photon-counting
detection have quite different drawbacks. The homodyne detection
is not sensitive to ``real" photons that are in modes unmatched
with the detected (local oscillator) mode, but it is quite
sensitive to vacuum modes which couple into this detected mode. On
the other hand, the photon-counting detection is not sensitive to
vacuum noise, but it will detect photons in any modes.
Correspondingly, two experimental parameters must be used~: an
{\it homodyne efficiency} parameter $\eta$, which measures the
overlap between the desired signal mode and the detected mode
\cite{GG} ; and a {\it modal purity} parameter $\xi$, which
characterizes which fraction of the detected photons are actually
in the desired signal mode \cite{KK}. In the simplest approach,
the homodyne efficiency can be modelized by  a lossy beamsplitter,
taking out desired correlated photons. On the other hand, the
modal purity $\xi$ in our experiment cannot be modelized by
another lossy beamsplitter, because a small value of $\xi$
corresponds to unwanted firings of the APD, for which a squeezed
vacuum is still measured at the homodyne detection port. More
precisely, the measured probability distribution for a quadrature
$x$ will be taken as $P(x) = \xi \; P_{cond}(x) + (1-\xi) \;
P_{uncond}(x)$, where $P_{cond}(x)$ and $P_{uncond}(x)$ are
respectively the conditioned and unconditioned probability
distributions, which depend on the values of $s$, $R$ and $\eta$.

It is then easy to determine values of the parameters  $\eta$ and
$\xi$ fitting the experimental data. The procedure to measure
$\eta$  is well established from squeezing experiments
\cite{laporta}, and it can be cross-checked by comparing the
classical parametric gain and the measured degree of squeezing.
The procedure to measure $\xi$ is less usual, and amounts to
evaluate how many unwanted photons make their way through the
spatial and spectral filters which are used on the photon counting
channel. Ultimately, this estimated value of $\xi$ must fit with
the observed conditional probability distribution, since $\eta$ is
independantly obtained from squeezing measurements.

Experimentally, this procedure turns out to be quite successful,
and for instance we have plotted on fig.~\ref{squeezcond} the
amplified and deamplified conditional probability distributions,
using as parameters the parametric gain $\exp(2s) = 2.36$, the
homodyne efficiency $\eta = 0.75$, and the modal purity parameter
$\xi = 0.7$. We note that the value of $s$ is evaluated from the
measured squeezing (see fig.~\ref{squeez}), while $\eta$ is
obtained as $\eta = \eta_T \eta_H^2 \eta_D$, where the overall
transmission $\eta_T = 0.94$, the mode-matching visibility $\eta_H
= 0.92$, and the detectors efficiency  $\eta_D = 0.945$ are
independantly measured. Finally, the modal purity $\xi$ is fitted
to the data, and cross-checked as the ratio between the expected
and actual APD counting rates.

\begin{figure}[t]
%\vskip 8cm
\center \includegraphics{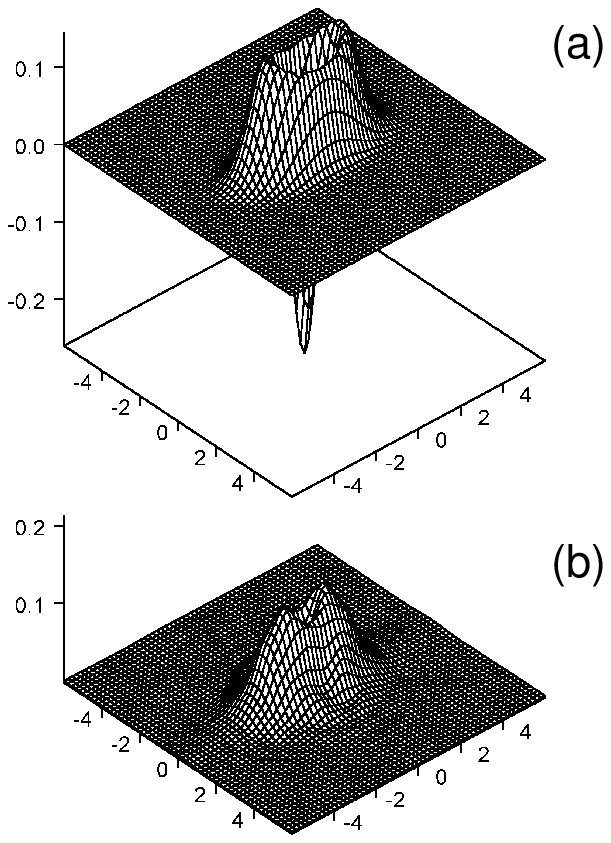} \caption{(a) Theoretical
Wigner function $W$ of the output state of the ``degaussification"
protocol, assuming $s=0.43$, $R=0.115$ and perfect detection
($\eta = \xi =1$). (b) Reconstructed Wigner function from the
experimental data ($\eta = 0.75$, $\xi =0.7$). The values of $W$
at the origin of phase space are respectively $W_{th}(0,0) =
-0.26$, and $W_{exp}(0,0) = 0.067$.} \label{wigner}
\end{figure}

In a last step, we have analysed our data using the standard
techniques of quantum tomography. We have recorded an histogram
with 40 bins for 6 different quadrature phase values $\theta$, and
about 5000 points for each histogram were acquired in a 3 hours
experimental run. The Wigner function displayed on
fig.~\ref{wigner} was then reconstructed using the Radon transform
\cite{ulf}, applied to the symetrized experimental data
$(P(x_\theta) + P(-x_\theta))/2$, without any correction for
measurement efficiency. It shows a clear dip at the origin, with a
central value of 0.067 while the maximum is at 0.12.

As usual, the conditions to get negative values of the measured
Wigner function are rather stringent, and require the presence of
a dip into the distribution probability associated to the squeezed
quadrature. Given our experimental parameters, this requires a
modal purity $\xi$ better than $0.85$, which was not
experimentally attainable while keeping the APD count rate above a
few tens per second. Nevertheless, we point out that by correcting
for the homodyne efficiency, the evaluated Wigner function of the
prepared state (just before homodyne detection) does assume a
negative value at the origin, $W_{cor}(0,0) \approx -0.06$.
Another interesting feature is that the non-gaussian dip on the
amplified quadrature is quite robust to losses and therefore can
be easily observed with our experimental parameters. This is
associated with a similarly robust ``squeezed volcano shape" of
the Wigner function.

We have described the first experimental observation of a
``degaussification" protocol, mapping individual femtosecond
pulses of squeezed light onto non-Gaussian states, by using only
linear optical elements and an avalanche photodiode. The observed
effect is closely related to the first step of an entanglement
distillation procedure for gaussian quantum continuous variables
\cite{eisert2}. This work should contribute to the future
development of quantum repeaters and long-range quantum
cryptography using continuous variables entanglement.

We thank F. Grosshans for his contribution to the early steps of
the experiment, and J. Fiur\'a\v{s}ek for useful comments. This
work was supported by the European IST/FET/QIPC program, and by
the French programs ``ACI Photonique" and ``ASTRE".

%%%%%%%%%%%%%%%%%%%%%%%%%%%%%%%%%%%%%%%%%%%%%%%%%%%%%%%%%%%%%%%%%%%%

%%%%%%%%%%%%%%%%%%%%%%%%%%%%%%%%%%%%%%%%%%%%%%%%%%%%%%%%%%%%%%%%%%%%%%%


\begin{references}


\bibitem{prl} F. Grosshans and Ph. Grangier,  Phys. Rev. Lett. {\bf 88}, 057902
(2002).

\bibitem{GVAWBCG03} F. Grosshans, G. Van Assche, J. Wenger, R. Brouri,
N.J. Cerf and Ph. Grangier, Nature {\bf421}, 238 (2003).

\bibitem{qic} F. Grosshans, N. J. Cerf, J.  Wenger, R. Tualle-Brouri
 and Ph. Grangier, Quant. Inf. Comput. {\bf 3}, 535 (2003).

\bibitem{gc} F. Grosshans and N.J. Cerf, Phys. Rev. Lett. {\bf 92}, 047905 (2004).

\bibitem{IVAC} S. Iblisdir, G. Van Assche and N.J. Cerf, e-print quant-ph/0312018.

\bibitem{gisin} N. Gisin, G. Ribordy, W. Tittel, and H. Zbinden,
Rev. Mod. Phys. {\bf 74}, 145 (2002).

\bibitem{qr} H.J. Briegel, W. Dur, J.I. Cirac and P. Zoller, Phys. Rev. Lett. {\bf 81}, 5932 (1998).

\bibitem{bennett} C.H. Bennett \etal, Phys. Rev. Lett. {\bf 70}, 1895 (1993).


\bibitem{eisert1} J. Eisert, S. Scheel and M.B. Plenio, Phys. Rev. Lett. {\bf 89}, 137903
(2002).

\bibitem{cirac} G. Giedke and J.I. Cirac, Phys. Rev. A {\bf 66}, 032316
(2002).

\bibitem{eisert2} D.E. Browne, J. Eisert, S. Scheel and M.B. Plenio, Phys. Rev. A {\bf 67},
062320 (2003).

\bibitem{laporta} R.E. Slusher, P. Grangier, A. LaPorta, B. Yurke
and M.J. Potasek, Phys. Rev. Lett. {\bf 59}, 2566 (1987).

\bibitem{Gid} A. LaPorta and R.E. Slusher, Phys. Rev. A {\bf 44}, 2013
(1991).

\bibitem{ek} A.I. Lvovsky \etal, Phys. Rev. Lett. {\bf 87}, 050402 (2001).

\bibitem{GG} F. Grosshans and P. Grangier, Eur. Phys. J. D. {\bf 14}, 119
(2001).

\bibitem{KK} T. Aichele, A.I. Lvovsky and S. Schiller, Eur. Phys. J. D. {\bf 18}, 237
(2002).

\bibitem{ulf} U. Leonhardt, \textit{Measuring the quantum state of
light} (Cambridge University Press, Cambridge, 1997).


\end{references}
\end{document}